\documentclass[11pt,twocolumn,english,american,aip,jap,numerical,reprint,groupedaddress]{revtex4-1}

\usepackage[T1]{fontenc}
\usepackage[latin9]{inputenc}
\usepackage{color}
\usepackage{babel}
\usepackage{float}
\usepackage{bm}
\usepackage{amsmath}
\usepackage{amssymb}
\usepackage{graphicx}
\usepackage{esint}
\usepackage[unicode=true,pdfusetitle,
 bookmarks=true,bookmarksnumbered=false,bookmarksopen=false,
 breaklinks=true,pdfborder={0 0 1},backref=false,colorlinks=true]
 {hyperref}
\usepackage{breakurl}

\makeatletter

\providecommand{\tabularnewline}{\\}

\@ifundefined{textcolor}{}
{%
 \definecolor{BLACK}{gray}{0}
 \definecolor{WHITE}{gray}{1}
 \definecolor{RED}{rgb}{1,0,0}
 \definecolor{GREEN}{rgb}{0,1,0}
 \definecolor{BLUE}{rgb}{0,0,1}
 \definecolor{CYAN}{cmyk}{1,0,0,0}
 \definecolor{MAGENTA}{cmyk}{0,1,0,0}
 \definecolor{YELLOW}{cmyk}{0,0,1,0}
}

\AtBeginDocument{
  
}

\makeatother

\begin{document}

\title{Effect of dipolar interactions and DC magnetic field on the specific
absorption rate of an array of magnetic nanoparticles}

\author{J.-L. Déjardin$^{1}$}

\email{dejardin@univ-perp.fr}

\author{F. Vernay$^{1}$}

\email{francois.vernay@univ-perp.fr}

\author{M. Respaud$^{2}$}

\email{marc.respaud@insa-toulouse.fr}

\author{H. Kachkachi$^{1}$}

\email{hamid.kachkachi@univ-perp.fr}

\affiliation{$^{1}$Laboratoire PROMES-CNRS (UPR-8521) \& Université de Perpignan
Via Domitia, Rambla de la thermodynamique, Tecnosud, 66100 Perpignan,
FRANCE.\linebreak{}
$^{2}$Laboratoire de Physique et Chimie des Nano-Objets, INSA, 135
Avenue de Rangueil, 31077 Toulouse, FRANCE }

\date{\today}
\begin{abstract}
We address the issue of inter-particle dipolar interactions in the
context of magnetic hyperthermia. More precisely, the main question
dealt with here is concerned with the conditions under which the specific
absorption rate is enhanced or reduced by dipolar interactions. For
this purpose, we propose a theory for the calculation of the AC susceptibility,
and thereby the specific absorption rate, for a monodisperse two-dimensional
assembly of nanoparticles with oriented anisotropy, in the presence
of a DC magnetic field, in addition to the AC magnetic field. We also
study the competition between the dipolar interactions and the DC
field, both in the transverse and longitudinal configurations. In
both cases, we find that the specific absorption rate has a maximum
at some critical DC field that depends on the inter-particle separation.
In the longitudinal setup, this critical field falls well within the
range of experiments.
\end{abstract}
\maketitle

\section{Introduction}

Today magnetic hyperthermia is one of the most promising applications
of magnetic nanoparticles. This is an experimental medical treatment
of cancer that has recently attracted numerous investigations from
the physics perspective. \citep{Carrey_JAP2011,Mehdaoui_AFM2011,Haase_Nowak_PRB85_2012,martinez2013learning,condeetal15jpcc,koslap2015nanotechrev,Ruta_ScientificReports_2015,Landi_JMMM_2017}
It consists in injecting in tumor cells magnetic nanoparticles
whose magnetization is then excited by an external AC magnetic field
into a fast switching motion. As a consequence, there is an elevation
of temperature of several Kelvins inside the cells that eventually
leads to their destruction. One of the most relevant quantities
to the efficiency of this process is what is called the \emph{specific
absorption rate} (SAR), which is defined as the power absorbed by
a magnetic sample subjected to an external AC field
\begin{equation}
{\rm SAR}=\mathfrak{Re}\left(\frac{\mu_{0}\omega}{2\pi}\oint_{{\rm cycle}}\bm{M}\cdot d\bm{H}_{{\rm ac}}\right).\label{eq:SAR_definition}
\end{equation}

The integration here is performed over one cycle of the magnetic field
and gives the energy dissipation per cycle. $\bm{M}$ is the sample's
magnetization and $\omega$ the angular frequency of the AC magnetic
field $\bm{H}_{AC}=H_{0}\exp\left(i\omega t\right)\bm{e_{x}}$. By
integration of Eq. (\ref{eq:SAR_definition}), it can be shown within
the framework of linear-response theory that the SAR is directly proportional
to the imaginary component of the AC susceptibility $\chi^{\prime\prime}\left(\omega\right)$
\citep{Hergt_etal_JPMC2006,Ahrentorp_etal_aipcp2010}. More precisely,
we have
\begin{equation}
{\rm SAR}=\frac{\mu_{0}\omega}{2\pi}H_{0}^{2}\chi^{\prime\prime}\left(\omega\right).\label{eq:SAR_chi_ac}
\end{equation}
See also Ref. \onlinecite{rosensweig02j3m} for a detailed derivation
of the related volumetric power dissipation.

Hence, computing the SAR for an assembly of nanoparticles can be achieved
upon obtaining its AC susceptibility. If we denote by $\chi_{\mathrm{eq}}$
the equilibrium susceptibility and by $\Gamma$ the relaxation rate
(the inverse longitudinal relaxation time $\tau=\Gamma^{-1}$), the
AC susceptibility may be computed according to the Debye model \citep{garpal00acp,azekac07prb}
\begin{equation}
\chi\left(\omega\right)=\frac{\chi^{\mathrm{eq}}}{1+i\omega\Gamma^{-1}}.\label{eq:DebyeXiac}
\end{equation}

Therefore, upon computing the equilibrium susceptibility $\chi^{\mathrm{eq}}$
and the relaxation rate $\Gamma$ of the assembly, in the presence
of dipolar interactions (DI) and a DC magnetic field, we can investigate
the effects of the latter two contributions on the SAR. This is the
main task of the present work. Accordingly, we will study the effects
of dipolar interactions and DC magnetic field on the SAR of a mono-disperse
assembly of magnetic nanoparticles arranged in a regular super-lattice.
Our main objective here is to investigate the conditions regarding
DI and DC field under which the SAR may be enhanced. Magnetic hyperthermia
makes use of a kind of ferrofluid, \emph{i.e.} an ensemble of (ferro)
magnetic nanoparticles floating in a fluid. In this work we consider
instead a solid matrix in which the nanoparticles are embedded and
spatially arranged. However, in order to investigate the qualitative
features of the SAR as a function of the assembly concentration and
magnetic DC field, we resort to a simple analytical formalism which
still captures the main behavior with respect to these two parameters.

The article is organized as follows: in the next Section we present
our model and hypotheses. Section \ref{sec:Susceptibility} is devoted
to the calculation of the AC susceptibility within the framework of
the Debye model. This requires the calculation of the equilibrium
susceptibility as well as the relaxation rate. We finally obtain an
expression of the SAR as a function of the assembly concentration
$C_{{\rm v}}$ and applied DC field. The article ends with our concluding
remarks and perspectives.

\section{\label{sec:Model-and-Hypotheses}Model and Hypotheses}

We consider a monodisperse assembly of $\mathcal{N}$ single-domain
nanoparticles with oriented (effective) uniaxial anisotropy, each
having a magnetic moment $\bm{m}_{i}=m_{i}{\bf s}_{i},\,i=1,\cdots,{\cal N}$
of magnitude $m$ and direction ${\bf s}_{i}$, with $\vert{\bf s}_{i}\vert=1$.
For the sake of simplicity, and without loss of generality, we focus
on a simple geometry: the assembly is organized into a simple cubic
two-dimensional super-lattice of parameter $a$. Indeed, the approach
adopted here can easily be extended to more general situations upon
computing the super-lattice sums of the corresponding geometry and
spatial configuration. This method is general and the 
standard sums involved in such calculations have already been introduced 
in similar contexts where the dipolar interactions have to be taken into account
\citep{jongar01prb,azzeggagh_EPJB2005}. Setting 
up the assembly in the $xy$ plane,
each nanoparticle of volume $V$ is attributed an (effective) uniaxial
anisotropy constant $K_{{\rm eff}}$ with an easy-axis in the $z$
direction. Indeed, even if the nanoparticles are modeled here as spheres,
we assume that asperities on their outer shell and related surface
effects may induce an effective easy axis for their resultant magnetic
moment. Furthermore, we assume that these easy axes are all pointing
in the $z$ direction. The present calculations can, of course, be
extended so as to include volume and anisotropy-easy axis distributions
using a fully numerical approach. However, as stated earlier, in this
work we would like to focus on the qualitative behavior of the SAR
in the presence of DI and a DC magnetic field and derive simple formulae
for practical use.

For later use we introduce the (dimensionless) anisotropy-energy barrier
$\sigma=K_{{\rm eff}}V/k_{B}T$. $K_{{\rm eff}}$ is considered to
be the largest energy scale of our model (\emph{i.e.} $\sigma\gg1$), meaning
that the anisotropy barrier is the dominant term in the expression of the energy.
This limit applies to most hyperthermia experiments (at temperature
$T\simeq318{\rm K}$) on iron-cobalt nanoparticles of volume $V\sim5.23\times10^{-25}{\rm m^{3}}$
(i.e. spheres of radius $R=5\ {\rm nm}$) with an effective anisotropy
constant $K_{{\rm eff}}\sim4.5\times10^{4}{\rm J.m^{-3}}$. Indeed,
in this case one has $\sigma\simeq5.4$. These assumptions also apply
to the various systems investigated in the literature \citep{Lacroix_etal_JAP2009,mehdaouietal12apl,Branquinho2013Effect-of-magne,kenya2013}. 

The energy of a magnetic moment $\bm{m}_{i}$ interacting with the
other moments of the assembly and subjected to an external DC magnetic
field $\bm{H}_{{\rm ex}}=H_{{\rm DC}}\bm{e_{z}}$, reads (after multiplying
by $-\beta\equiv-1/k_{B}T$)

\begin{equation}
\mathcal{E}_{i}={\cal E}_{i}^{(0)}+{\cal E}_{i}^{\mathrm{DI}},\label{eq:DDIAssemblyEnergy}
\end{equation}
where 
\begin{equation}
{\cal E}_{i}^{(0)}=x\,{\bf s}_{i}\cdot\bm{e_{z}}+\sigma\left(\mathbf{s}_{i}\cdot\bm{e_{z}}\right)^{2}\label{eq:Energy-SingleNP}
\end{equation}
 is the energy of a single (noninteracting) nanoparticle located at
site $i$. This includes the Zeeman and anisotropy terms, with $x=\beta mH_{{\rm DC}}$.
The second term in Eq. (\ref{eq:DDIAssemblyEnergy}) is the contribution
from the long-range DI, 
\begin{equation}
{\cal E}_{i}^{\mathrm{DI}}=\xi\sum_{j<i}{\bf s}_{i}\cdot{\cal D}_{ij}\cdot{\bf s}_{j},\label{eq:Energy-DIcontribution}
\end{equation}
with the usual tensor 
\begin{equation}
{\cal D}_{ij}\equiv\frac{1}{r_{ij}^{3}}\left(3\bm{e}_{ij}\bm{e}_{ij}-1\right)\label{eq:TensorDI}
\end{equation}
where $\bm{r}_{ij}=\bm{r}_{i}-\bm{r}_{j}$, $r_{ij}=\left|\bm{r}_{ij}\right|$and
$\bm{e}_{ij}=\bm{r}_{ij}/r_{ij}$ a unit vector along the link $i\rightarrow j$.
In dimensionless units the DI coefficient $\xi$ reads 
\begin{equation}
\xi=\left(\frac{\mu_{0}}{4\pi}\right)\left(\frac{m^{2}/a^{3}}{k_{B}T}\right).\label{eq:DI-xi}
\end{equation}

Alternatively, the DI can be expressed as the result of the DI field
$\bm{\Xi}_{i}$ acting on $\bm{m}_{i}$ with
\begin{equation}
\bm{\Xi}_{i}=\xi\sum_{j\ne i}\mathcal{D}_{ij}\cdot\mathbf{s}_{j}.\label{ZetaEffectiveField}
\end{equation}
The main purpose of our investigation is to derive (semi)-analytical
formulae that account for the effect of DI and DC field on the SAR.
Accordingly, we limit the present study to low particle concentrations
(\emph{i.e.} $\xi\ll1$) since then we can use perturbation theory
to investigate the behavior of the SAR upon varying $\xi$ and $x$. 

Now, a word is in order regarding Debye's formula (\ref{eq:DebyeXiac}).
In Ref. \onlinecite{jongar01epl} the contribution of DI to the relaxation
rate $\Gamma$ was obtained in the adiabatic approximation. More precisely,
for an ensemble of weakly coupled magnetic moments, one assumes that
there are mainly two time scales: 
\begin{itemize}
\item the \textquotedblleft single-particle\textquotedblright{} time scale
$\tau_{s}\sim1/\gamma H_{K}$, where $\gamma$ is the gyromagnetic
factor and $H_{K}$ the anisotropy field of the particle. $\tau_{s}$
is the (intrinsic) characteristic time of the dynamics of an individual
magnetic moment, 
\item the \textquotedblleft collective\textquotedblright{} time scale $\tau=\Gamma^{-1}$
that corresponds to the dynamics of the \textquotedblleft soft\textquotedblright{}
collective state induced by (weak) DI in the whole assembly. 
\end{itemize}
Thus, in the adiabatic approximation one assumes that $\tau\gg\tau_{s}$,
which means that because of the weak DI, what is happening at the
level of individual moments is conveyed with delay to the other moments
and eventually to the whole assembly. Equivalently, this implies that
when the dynamics of an individual magnetic moment is probed and the
relaxation rate is being measured, one assumes that the other moments
are \textquotedblleft frozen in time\textquotedblright{} and exert
only a static \textquotedblleft molecular\textquotedblright{} field
on the moment considered. Hence, the latter is subject to a static
effective field due to DI, in addition of course to the anisotropy
and Zeeman fields. In conclusion, in the present approach, it is understood
that the collective dynamics of the system is assumed to be dominated
by a \textquotedblleft slow\textquotedblright{} mode corresponding
to one (longitudinal) relaxation time. This comes out as a correction
to the Debye formula and which is taken into account in the present
work by a DI correction of the equilibrium susceptibility and the
longitudinal relaxation time. This is done in analogy with various
works on the extensions of Debye's model in the context of dielectric
relaxation. Indeed, the simplest model of orientational relaxation
is that of rotational diffusion first proposed by Debye \citep{debye1929Dover}
in which rigid molecules diffuse independently. Zwanzig\citep{zwanzig_jcp63p2766}
{[}see also Ref. \onlinecite{Berne_jcp62p1154}{]} later investigated
how DI affect this model on a rigid cubic lattice. It was shown that,
to first-order in concentration (or DI coefficient $\xi$), one obtains
only one (longitudinal) relaxation time shifted from the molecular
relaxation time by some correction factor that depends on the density
of the lattice, with a very good agreement with Debye's formula. The
correction factor vanishes at vanishing density. The high-order corrections
to the Debye formula are responsible for new relaxation times that
become relevant at much higher frequencies, \emph{e.g.} in FMR measurements.

\section{AC Susceptibility \label{sec:Susceptibility} }

AC susceptibility can be written as $\chi\left(x,\sigma,\xi,\omega\right)=\chi^{\prime}-i\chi^{\prime\prime}$
with its real real and imaginary components given by
\begin{eqnarray}
\chi^{\prime} & = & \chi^{\mathrm{eq}}\frac{1}{1+\eta^{2}},\ \chi^{\prime\prime}=\chi^{\mathrm{eq}}\frac{\eta}{1+\eta^{2}},\label{eq:ReImXiAssembly}
\end{eqnarray}
with $\chi^{\mathrm{eq}}$ being the equilibrium susceptibility, \emph{i.e.} 
the response of the magnetic system to the static
magnetic DC field $\bm{H}_{\rm{ex}}$. On the other hand, the AC susceptibility 
is the response of the system to $\bm{H}_{AC}$, the AC magnetic field. 
In the present work, we remain within the linear-response regime since $H_0$
is assumed to be too small to change the energy states of the magnetic system. For this
reason,  $\bm{H}_{AC}$ does not need to be included in the Hamiltonian 
that is used for determining the equilibrium states of the system.
The parameter $\eta$ in Eq. (\ref{eq:ReImXiAssembly}) is given by 
$\eta=\omega\Gamma^{-1}$ where $\Gamma$ is the relaxation rate associated
with the magnetization switching between its minimal-energy orientations.

\begin{figure}[H]
\begin{centering}
\includegraphics[width=0.95\columnwidth]{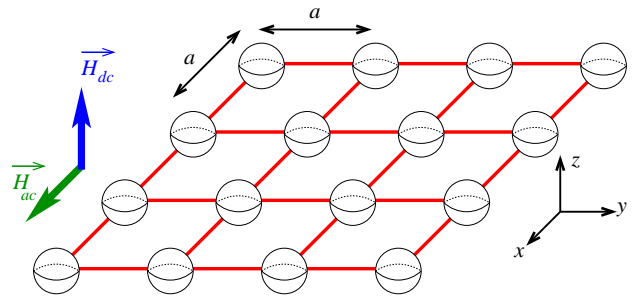}
\par\end{centering}

\centering{}\caption{\label{fig:Oblate-2D-sample}2D assembly of nano-spheres on a square
super-lattice of parameter $a$. The external DC field is applied
along the $z$-axis, the AC field lies within the $xy$-plane. 
We assume the assembly to be monodisperse with
all anisotropy easy axes oriented in the $z$ direction.}
\end{figure}

In the presence of (weak) DI, in Refs. \onlinecite{azzeggagh_EPJB2005, sabsabietal13prb},
$\chi^{\mathrm{eq}}$ was shown to be given by 
\begin{equation}
\chi^{\mathrm{eq}}\simeq\chi_{\mathrm{free}}^{\mathrm{eq}}+\tilde{\xi}\chi_{\mathrm{int}}^{\mathrm{eq}},\label{eq:SuscepEq}
\end{equation}
where $\tilde{\xi}\equiv\xi\mathcal{C}^{\left(0,0\right)}$, with
$\mathcal{C}^{\left(0,0\right)}$ being a lattice sum that can be
expressed in terms of the assembly demagnetizing factor along $z$
{[}see Section \ref{sec:Lattice-sums}{]}. For instance, for the specific
case of a square sample, as shown in Fig. \ref{fig:Oblate-2D-sample},
the lattice sum is given by (in the thermodynamic limit) $\mathcal{C}^{\left(0,0\right)}\simeq-9$,
whereas it is positive for a prolate sample of the form $L\times L\times2L$
for which $\mathcal{C}^{\left(0,0\right)}\simeq1.7$.

$\chi_{{\rm free}}^{{\rm eq}}$ and $\chi_{{\rm int}}^{{\rm eq}}$
represent the contributions to the linear equilibrium susceptibility
without DI and with DI, respectively. Their explicit derivation can
be found in Refs. \onlinecite{jongar01prb, sabsabietal13prb}. Here
we only report the main result for the longitudinal DC field case 
\begin{eqnarray}
\chi_{\mathrm{free}}^{\mathrm{eq}} & = & \frac{\mu_{0}m^{2}}{k_{B}T}\left[1-\frac{1}{\sigma}-\left(1-\frac{2}{\sigma}\right)x^{2}\right],\label{eq:XiEqFree}\\
\nonumber \\
\chi_{\mathrm{int}}^{\mathrm{eq}} & = & \frac{\mu_{0}m^{2}}{k_{B}T}\left[1-\frac{2}{\sigma}-4\left(1-\frac{3}{\sigma}\right)x^{2}\right].\label{eq:XiEqIntContr-1}
\end{eqnarray}
Note that these two expressions are valid for $x\lesssim0.5$. For
larger values of $x$, we must use the expressions given in Eqs. (3.85)
\& (3.39) of Ref. \onlinecite{garpal00acp}, with $h=x/2\sigma$. In the present notations,
these are rewritten as follows

\selectlanguage{english}%
\begin{widetext}

\selectlanguage{american}%
\begin{eqnarray}
\chi_{\mathrm{free,GP}}^{\mathrm{eq}} & \simeq & \frac{\mu_{0}m^{2}}{k_{B}T}\frac{1}{\left(\cosh x-h\sinh x\right)^{2}}\nonumber \\
 & \times & \left\{ \left(1-h^{2}\right)-\frac{1}{\sigma}+\frac{1}{8\sigma^{2}}\left[1-\frac{\left(1+6h^{2}+h^{4}\right)\cosh\left(2x\right)-4h\left(1+h^{2}\right)\sinh\left(2x\right)}{\left(1-h^{2}\right)^{2}}\right]\right\} ,\label{eq:Xifree GP}\\
\nonumber \\
\chi_{\mathrm{int,GP}}^{\mathrm{eq}} & \simeq & \frac{\mu_{0}m^{2}}{k_{B}T}\frac{1}{2}\frac{\partial^{2}}{\partial x^{2}}\left\{ \tanh x\left[1-\frac{1}{2\sigma}\left(1+\frac{2x}{\sinh\left(2x\right)}\right)-\frac{1}{8\sigma^{2}}\left(4-x\frac{\sinh\left(2x\right)-2x}{\cosh^{2}x}\right)\right]\right\} ^{2}.\label{eq:Xi int GP}
\end{eqnarray}

\selectlanguage{english}%
\end{widetext}

\begin{figure}
\begin{centering}
\includegraphics[width=0.9\columnwidth]{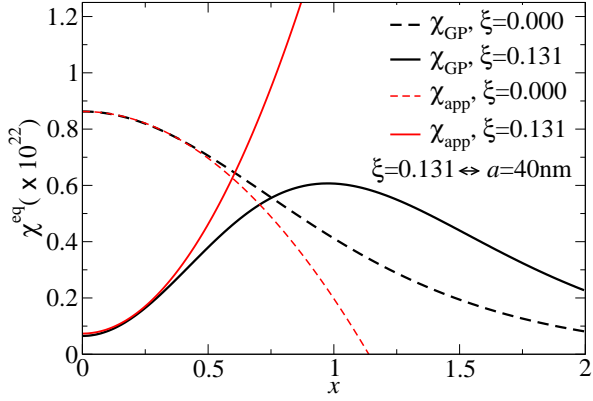}
\par\end{centering}

\caption{Linear susceptibility $\chi^{\mathrm{eq}}$ as a function of the (reduced)
longitudinal DC field $x$ in the absence of DI (dashed lines $\xi=0$) and in
the presence of DI (continuous lines $\xi=0.131$, \emph{i.e.} with
super-lattice parameter $a=40$ nm). The curves in black are plots
of Eqs. (\ref{eq:Xifree GP},\ref{eq:Xi int GP}), of Ref. \onlinecite{garpal00acp}
(GP stands for Garcia-Palacios), while the curves in red are plots
of the expressions in Eqs. (\ref{eq:XiEqFree},\ref{eq:XiEqIntContr-1}).\label{fig:Linear-susceptibility-ksi=00003D0 and ksi non nul}}
\end{figure}

\selectlanguage{american}%
The difference between $\chi^{{\rm eq}}$ in Eqs. (\ref{eq:XiEqFree},
\ref{eq:XiEqIntContr-1}) and Eqs. (\ref{eq:Xifree GP}, \ref{eq:Xi int GP})
is shown in Fig. \ref{fig:Linear-susceptibility-ksi=00003D0 and ksi non nul}.
Their comparison allows us to establish the validity of the approximate
expressions in Eqs. (\ref{eq:XiEqFree}, \ref{eq:XiEqIntContr-1}). 
Note in passing that all these expressions are only valid in the limit
$\sigma\ll 1$, which is relevant for the specific case of hyperthermia. 
More general expressions can be obtained if one rederives the equilibrium 
susceptibility from exact expressions of the magnetization as given in 
Refs. \onlinecite{Bakuzis_jmmm_2001,Bakuzis_2017arXiv170202022C}. We have
checked that the $\frac{1}{\sigma}$-series expansion of the latter gives 
the same expressions as used here. 
The main feature that appears in the presence of DI is clearly shown
by the black curves in Fig. \ref{fig:Linear-susceptibility-ksi=00003D0 and ksi non nul}:
the competition between on one hand, the DI that tend to maintain
the magnetization within the $xy$-plane, and on the other, the external
DC field together with anisotropy that tend to align the magnetic
moments along the $z$-direction, leads to a nonmonotonic behavior
of $\chi^{{\rm eq}}$ with a maximum at an external DC field $x_{{\rm m}}\sim0.9$.
In the limit of high anisotropy-energy barrier, namely $\sigma\gg1$,
this maximum can be analytically obtained; it only depends on the
DI parameter $\tilde{\xi}$ as follows
\begin{equation}
x_{{\rm m}}={\rm Arcsech}\left[\frac{1}{\sqrt{3}}\sqrt{1-\frac{1}{2\tilde{\xi}}}\right].\label{eq:field_max_chieq}
\end{equation}
For the present case, with $\xi=0.131$ and lattice sum $\mathcal{C}^{\left(0,0\right)}=-9$
(\emph{i. e.} $\tilde{\xi}=-0.972$), we obtain $x_{{\rm m}}\simeq0.875$
which is in agreement with the result in Fig. \ref{fig:Linear-susceptibility-ksi=00003D0 and ksi non nul}.

The remaining task in the calculation of the AC susceptibility (\ref{eq:ReImXiAssembly})
is to compute the relaxation rate, or the (dimensionless) relaxation
time $\eta$. For a single (noninteracting) particle the relaxation
rate $\Gamma_{0}$, in a longitudinal DC field, is given by the Néel-Brown
formula \citep{aha69pr}. The longitudinal relaxation rate given by Néel-Brown
formula is valid for any $\sigma$ as long as the two-well character of the
energy potential is preserved. Indeed, the relaxation rate computed within
Brown's or Langer's approach is  based on the notion of escape rate (or
first-passage time) 
from a metastable minimum to a more stable minimum, through a saddle
point. Hence, the existence of the saddle point and of the minima has to be
well defined in the energy landscape.  Therefore, the Néel-Brown expression 
for the relaxation rate is valid for $\sigma$ ranging from a few units to a
few tens, which is the case for the magnetic systems used in hyperthermia
applications. The Néel-Brown formula reads  
\begin{equation}
\begin{array}{lll}
\tau_{D}\Gamma_{0} & = & \dfrac{\sigma^{1/2}\left(1-h^{2}\right)}{\sqrt{\pi}}\\
 &  & \times\left[\left(1+h\right)e^{-\sigma\left(1+h\right)^{2}}+\left(1-h\right)e^{-\sigma\left(1-h\right)^{2}}\right],
\end{array}\label{eq:RRFreeAssembly}
\end{equation}
where $h=x/2\sigma$ and $\tau_{D}\sim2\times10^{-10}-2\times10^{-12}{\rm s}$
is the free-diffusion time. 

In the presence of (weak) DI, Jönsson and Garcia-Palacios \citep{jongar01epl}
showed that the relaxation rate depends on the damping factor $\lambda$
and is expressed as a quadratic function of the longitudinal and transverse
components of the DI field $\bm{\Xi}_{i}$. Accordingly, the explicit
expression of $\eta$ depends on the lattice through sums such as
$\mathcal{R}=2\sum_{j\neq i}r_{ij}^{-6}$, $\mathcal{T}=\sum_{j\neq i}\left(\bm{e}\cdot{\cal D}_{ij}\bm{e}\right)^{2}$
\citep{jongar01epl,Vernay_etal_acsucept_PRB2014}. More precisely,
we have 
\begin{equation}
\eta=\frac{\omega}{\Gamma}=\frac{\omega}{\Gamma_{0}}\left[1-\frac{\xi^{2}}{6}\mathcal{S}\left(\lambda\right)\right],\label{eq:eta_inverserelaxation_rate}
\end{equation}
with $\mathcal{S}\left(\lambda\right)=\left(1+F\left(\lambda\right)\right)\mathcal{R}+\left(3\mathcal{T}-\mathcal{R}\right)\left(1-F\left(\lambda\right)/2\right)S_{2}$.
The function $F(\alpha)$ is given by \citep{garetal99pre} 
\begin{equation}
F(\alpha)=1+2(2\alpha^{2}e)^{1/(2\alpha^{2})}\gamma(1+\dfrac{1}{2\alpha^{2}},\dfrac{1}{2\alpha^{2}}),\label{eq:DampingFunction}
\end{equation}
where $\gamma(a,z)=\int_{0}^{z}dt\,t^{a-1}e^{-t}$ and $\alpha=\lambda\sqrt{\sigma}$. 

Finally, substituting in Eq. (\ref{eq:ReImXiAssembly}) the expressions
(\ref{eq:XiEqFree}, \ref{eq:XiEqIntContr-1}) or (\ref{eq:Xifree GP},\ref{eq:Xi int GP})
for $\chi^{\mathrm{eq}}$ and (\ref{eq:RRFreeAssembly}, \ref{eq:eta_inverserelaxation_rate})
for $\Gamma$, renders an expression of the AC susceptibility for
the assembly in the presence of DI. Therefore, the DI contribute to
$\chi^{\prime\prime}$ through the relaxation rate as well as the
equilibrium susceptibility. However, it can easily be seen that for
low concentrations the DI correction is mostly brought in by the equilibrium
susceptibility. Hence, to first order in $\xi$ we can write ($\eta_{0}=\omega\Gamma_{0}^{-1}$)\citep{Vernay_etal_acsucept_PRB2014}
\begin{equation}
\chi^{\prime\prime}\simeq\frac{\eta_{0}}{1+\eta_{0}^{2}}\left[\chi_{\mathrm{free}}^{\mathrm{eq}}+\tilde{\xi}\chi_{\mathrm{int}}^{\mathrm{eq}}\right].\label{eq:Chiseconde}
\end{equation}

\begin{figure}
\begin{centering}
\includegraphics[width=0.9\columnwidth]{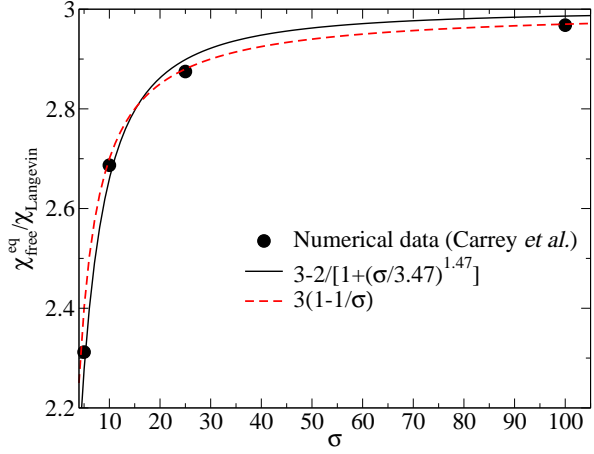}
\par\end{centering}

\caption{\label{fig:chi_chiLangevin_Ratio}$\chi_{{\rm free}}^{{\rm eq}}/\chi_{{\rm Langevin}}$
as a function of the anisotropy parameter $\sigma$. The numerical
data are extracted from Ref. \onlinecite{Carrey_JAP2011} and are compared
to the phenomenological expression $\left(3-\frac{2}{1+(\sigma/3.4)^{1.47}}\right)$
and to the analytical expression $3(1-1/\sigma)$.}

\end{figure}

Note that this result agrees with Eq. (34) of Ref. \onlinecite{Carrey_JAP2011}
for $\xi=0$. Indeed, the expression of $\chi_{\mathrm{free}}^{\mathrm{eq}}$
given in Eq. (\ref{eq:XiEqFree}) provides a clear basis for the phenomenological
formula of Eq. (38) given by Carrey \emph{et al}. In fact, the factor
$\frac{1}{3}\left(3-\frac{2}{1+(\sigma/3.4)^{1.47}}\right)$ with
\emph{ad-hoc} exponents and coefficients is extracted from a fitting
of the ratio $\chi_{{\rm free}}^{{\rm eq}}/\chi_{{\rm Langevin}}$,
supposedly with the aim to obtain an interpolation between the two
regimes $\sigma\ll1$ and $\sigma\gg1$. However, as we have already
mentioned, for applications to hyperthermia we have the typical values
of $\sigma\simeq5-30$. In this case, analytical calculations show
that it is a good approximation to replace the ratio $\chi_{{\rm free}}^{{\rm eq}}/\chi_{{\rm Langevin}}$
by $(1-1/\sigma)$, as can be seen in the square brackets in Eq. (\ref{eq:XiEqFree}).
This is clearly illustrated by the results in Fig. \ref{fig:chi_chiLangevin_Ratio}.

The DI parameter $\xi$ defined in (\ref{eq:DI-xi}) can be rewritten
in terms of the particles concentration $C_{{\rm v}}$ as \citep{sabsabietal13prb}
\begin{equation}
\xi=\frac{\mu_{0}}{4\pi}\frac{m^{2}}{k_{B}T}\frac{C_{{\rm v}}}{V}.\label{eq:XivsCV}
\end{equation}

\section{Specific Absorption Rate -- Effect of DC field and concentration\label{sec:SAR_Cv}}

\subsection{SAR for ``noninteracting assemblies''}

Let us first examine the behavior of the SAR as a function of the
applied DC field for noninteracting (\emph{i.e.} free) particles.
Within this approximation the SAR can be written as 
\begin{equation}
{\rm SAR}=\left(\frac{\mu_{0}}{2\pi}\right)\frac{\Gamma_{0}\eta_{0}^{2}}{1+\eta_{0}^{2}}H_{0}^{2}\chi_{\mathrm{free}}^{\mathrm{eq}}.\label{eq:SAR_free}
\end{equation}

In the case of a longitudinal ($\parallel$) DC field, the expressions
of $\Gamma_{0}$ and $\chi_{{\rm free}}^{\mathrm{eq}}$ are given
in Eqs. (\ref{eq:RRFreeAssembly}) and (\ref{eq:XiEqFree}) or (\ref{eq:Xifree GP}),
respectively. For a transverse ($\perp$) DC field, the general expression
of Eq. (\ref{eq:SAR_free}) still holds, one should simply replace
the expression of the relaxation rate $\Gamma_{0}$ and that of the
free susceptibility by the appropriate expressions \citep{Coffey_etalII_PRB_1998,garpal00acp},
namely 
\begin{equation}
\begin{array}{lll}
\tau_{D}\Gamma_{0}^{\perp} & = & \frac{\left[1-2h+\sqrt{1+4\lambda^{-2}h\left(1-h\right)}\right]\sqrt{1+h}}{2\pi\sqrt{h}}e^{-\sigma\left(1-h\right)^{2}},\\
\\
\chi_{{\rm free}}^{\mathrm{eq},\perp} & = & \left(\frac{\mu_{0}m^{2}}{k_{B}T}\right)\frac{1}{2\sigma}\frac{\left(1+h^{2}\right)\cosh x-2h\sinh x}{\left(1-h^{2}\right)\left(\cosh x-h\sinh x\right)}.
\end{array}\label{eq:DCField-Transverse}
\end{equation}

\begin{figure}
\begin{centering}
\includegraphics[width=0.95\columnwidth]{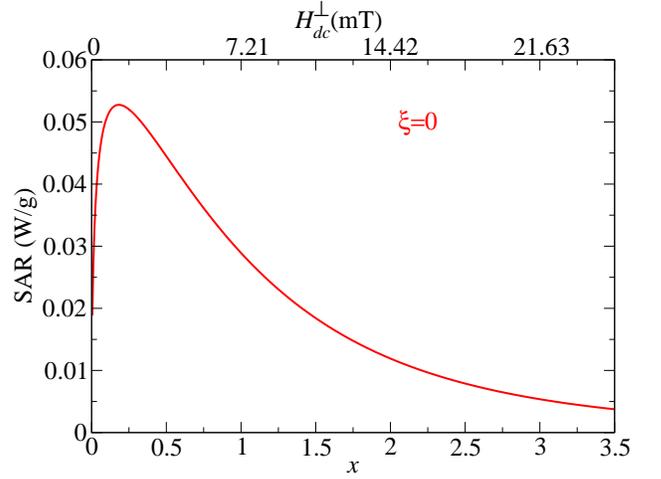}
\par\end{centering}

\caption{\label{fig:Plot-of-the SAR in trans field}SAR as a function of the
transverse DC magnetic field $H_{{\rm DC}}$ for $H_{0}=7.3\;\mathrm{mT,\;\omega=35.18\times10^{4}\;\mathrm{rad/s}}$. }

\end{figure}

Using these expressions one can see that the transverse susceptibility
$\chi_{{\rm free}}^{\mathrm{eq},\perp}$ only weakly depends on the
DC field $x$; only 3\% of change over the range $x=0-2$. On the
contrary, $\chi_{{\rm free}}^{\mathrm{eq},\parallel}$ changes by
an order of magnitude over the same range, as shown by the black dashed
curve in Fig. \ref{fig:Linear-susceptibility-ksi=00003D0 and ksi non nul}.
This explains why the effect of the DC field on the SAR is more pronounced
in the longitudinal case. The effect of a transverse DC field on the
SAR of non-interacting assemblies is shown in Fig. \ref{fig:Plot-of-the SAR in trans field}.
Notice the scale on the vertical axis.

The results are presented for external DC fields below 25 mT. Indeed,
we have to restrict ourselves to $H_{dc}^{\perp}<H_{K}=\frac{2K_{{\rm eff}}}{M_{s}}\simeq77\ {\rm mT}$
in order to preserve the picture of a two-well potential energy on
which the over-barrier escape rate theory\citep{lan68prl,brown79ieee}
is based. 

In the sequel, we will focus on the longitudinal geometry where more
pronounced effects are observed. This will be done in the presence
of DI.

\subsection{SAR for (weakly) interacting assemblies}

In the absence of a DC magnetic field (\emph{i.e.} $x=0$), we can
use Eqs. (\ref{eq:XivsCV}, \ref{eq:Chiseconde}) together with Eq.
(\ref{eq:SAR_chi_ac}) to write a relatively simple expression for
the SAR of a (weakly) interacting assembly in terms of the particles
concentration $C_{{\rm v}}$ 
\begin{equation}
\begin{array}{lll}
{\rm SAR} & = & \left(\frac{\mu_{0}}{2\pi}\right)\frac{\Gamma_{0}\eta_{0}^{2}}{1+\eta_{0}^{2}}\left(\frac{\mu_{0}m^{2}}{k_{B}T}\right)H_{0}^{2}\left[\left(1-\frac{1}{\sigma}\right)\right.\\
\\
 &  & +\left.\mathcal{C}^{\left(0,0\right)}\xi\left(C_{{\rm v}}\right)\left(1-\frac{2}{\sigma}\right)\right].
\end{array}\label{eq:SAR-x0xi}
\end{equation}

In particular, this expression shows that, to $1^{\mathrm{st}}$ order,
the SAR is linear in the concentration of the assembly with a strong
dependence on its shape via the coefficient $\mathcal{C}^{\left(0,0\right)}$
{[}see discussion in Section \ref{sec:Lattice-sums}{]}.

We now consider the evolution of the SAR as a function of the DI parameter
$\tilde{\xi}$, in a variable external DC magnetic field for which
we introduce the new parameter $h=x/2\sigma$. For numerical estimates,
we have used the physical parameters of ${\rm FeCo}$, the magnetic
material studied in Ref. \onlinecite{Lacroix_etal_JAP2009}, namely $M_{s}=1.162\times10^{6}\ {\rm A/m}$,
$K_{{\rm eff}}=4.5\times10^{4}{\rm J/m^{3}}$ with a density $\rho\simeq8300\ {\rm kg.m^{-3}}$.
In SI units the SAR is expressed in Watt per particle but it is more
commonly measured in W/g. We have also assumed that each nanoparticle
is a sphere of radius $R=5\ {\rm nm}$. For such a size but elongated
shape, which leads to a strong effective anisotropy, the blocking
temperature of an individual particle is $\sim60{\rm K}$. However,
the working temperature relevant to hyperthermia applications is $T=318\ {\rm K}$,
leading to the reduced anisotropy-energy barrier $\sigma\simeq5.4$.
This implies that the individual nanoparticles are in the superparamagnetic
state. This is an additional reason for which the SAR behavior is
mostly dictated by the equilibrium susceptibility, as stressed earlier. 

Regarding the AC magnetic field, we have set it at a small amplitude
$H_{0}=7.3\ {\rm mT}$ so as to remain in the linear regime and to
preserve the validity of the approach leading to Eq. (\ref{eq:SAR_chi_ac})
for the SAR. Indeed, we note that this amplitude is smaller than the
usual experimental value,\emph{ e. g.} 23 mT, as in Ref. \onlinecite{mehdaouietal12apl}.
As a consequence, since the SAR scales like $H_{0}^{2}$, the computed
value in this paper should be at least one order of magnitude lower
than that observed in experiments. In fact, as stressed earlier, our
aim here is not to achieve a quantitative agreement with experiments
regarding the SAR but rather to explain the role of DI and its possible
competition with an external DC magnetic field. An extension of the
present approach beyond the linear regime should be possible on the
basis of the developments in Refs. \onlinecite{zwanzig_jcp63p2766,Berne_jcp62p1154, garpalgar04prb}.

\begin{figure}
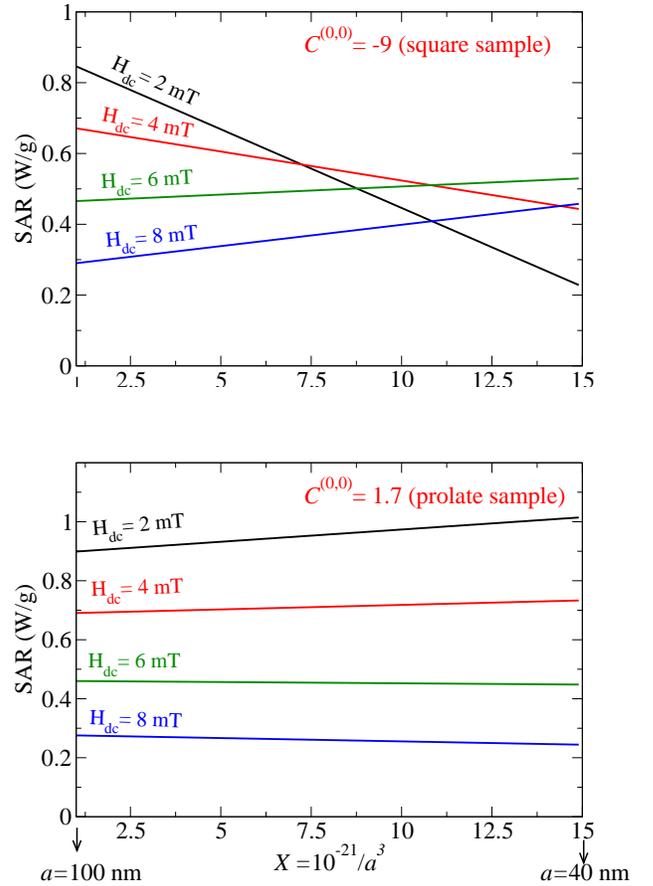

\begin{centering}
\includegraphics[width=0.95\columnwidth]{Fig5a.eps}
\par\end{centering}

$\ $

\begin{centering}
\includegraphics[width=0.95\columnwidth]{Fig5b.eps}
\par\end{centering}

\caption{\label{fig:SAR GP}SAR as a function of $X$ (see text) for various
values of the longitudinal field $H_{\mathrm{DC}}$ with $H_{0}=7.3\;\mathrm{mT,\;\omega=35.18\times10^{4}\;\mathrm{rad/s}}$.}
\end{figure}

Then, using Eqs. (\ref{eq:Xifree GP},\ref{eq:Xi int GP}) we obtain
the expression for the SAR of the now (weakly) interacting assembly
\begin{equation}
{\rm SAR}=\frac{\omega^{2}\Gamma_{0}\left(h\right)}{\omega^{2}+\Gamma_{0}^{2}\left(h\right)}\frac{\mu_{0}H_{0}^{2}}{2\pi}\left[\chi_{\mathrm{free},\mathrm{GP}}^{\mathrm{eq}}+\tilde{\xi}\chi_{\mathrm{int,\mathrm{GP}}}^{\mathrm{eq}}\right],\label{eq:SAR-GP}
\end{equation}
where the relaxation rate depends on the external DC field according
to Eq. (\ref{eq:RRFreeAssembly}). Expression (\ref{eq:SAR-GP}) is
plotted in Fig. \ref{fig:SAR GP} against the variable $X=10^{-21}/a^{3}$,
where $a$ is expressed in meters. 

\begin{figure}
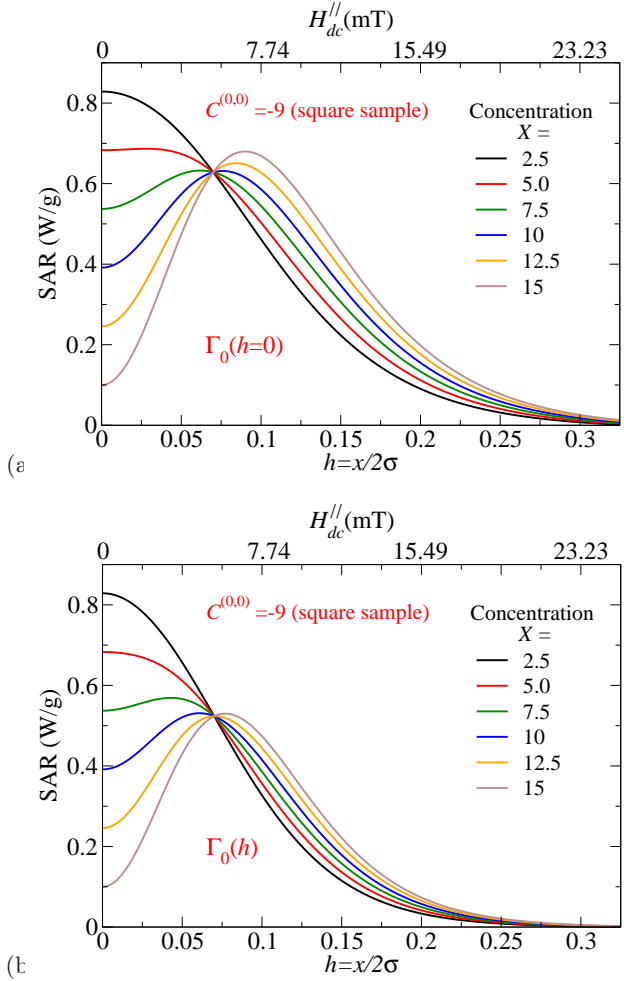

\begin{centering}
(a)\includegraphics[width=0.9\columnwidth]{Fig6a.eps}
\par\end{centering}

$\ $

\begin{centering}
(b)\includegraphics[width=0.9\columnwidth]{Fig6b.eps}
\par\end{centering}

\caption{\label{fig:SAR-of-h}The SAR as a function of the longitudinal DC field for various
values of the concentration $X$. (a) SAR obtained by setting $h=0$
in $\Gamma_{0}$ in Eq. (\ref{eq:SAR-GP}), (b) the SAR obtained from
Eq. (\ref{eq:SAR-GP}) for non zero field.}

\end{figure}

An interesting behavior is observed for oblate samples, as can be
seen in the upper panel in Fig. \ref{fig:SAR GP}, in the specific
case of a two-dimensional square sample (depicted in Fig. \ref{fig:Oblate-2D-sample}).
Here, the nano-spheres are placed on the vertices of a square lattice
lying in the $xy$-plane, with their effective anisotropy axes parallel
to the $z$-axis and the DC field applied parallel to the latter.
In the present setup, there is a competition between the DI and the
DC field. Consequently, the results in Figs. \ref{fig:SAR GP} and
\ref{fig:SAR-of-h} show that there are two ways to enhance the SAR,
namely by tuning either the concentration or the magnitude of the
DC field.

The curve crossing seen in the upper panel in Fig. \ref{fig:SAR GP}
can be understood upon analyzing the behavior of $\chi^{{\rm eq}}$
in the presence of DI, as shown in Fig. \ref{fig:Linear-susceptibility-ksi=00003D0 and ksi non nul}.
Indeed, one clearly sees that for an inter-particle distance of $40\ {\rm nm}$
\emph{i.e.} $X\sim15$, the equilibrium susceptibility reaches a maximum
for a DC field of about $H_{{\rm DC}}^{{\rm max}}=\frac{k_{B}T}{m}x_{{\rm max}}=\frac{k_{B}T}{m}0.8\simeq6\ {\rm mT}$.
According to Eq. (\ref{eq:SAR-GP}), this means that the SAR should
exhibit a maximum around this field. 
 However, this interpretation should take into account the fact that
the SAR in Eq. (\ref{eq:SAR-GP}) depends on $h$ not only through
the equilibrium susceptibility but also through the relaxation rate
$\Gamma_{0}\left(h\right)$. In order to clarify the effect of the
dynamics, let us now further focus on the effect of the DC field on
the SAR. For this purpose, in Fig. \ref{fig:SAR-of-h} we plot the
SAR as a function of the field, as rendered by Eq. (\ref{eq:SAR-GP}).
In Fig. \ref{fig:SAR-of-h} (a), we plot Eq. (\ref{eq:SAR-GP}) where
we have used the relaxation rate in zero field, \emph{i.e.} $\Gamma_{0}\left(h=0\right)$,
which means that we simply adopt the Arrhenius law for the relaxation
rate $\Gamma_{0}$. On the other hand, in Fig. \ref{fig:SAR-of-h}
(b) we plot the full expression (\ref{eq:SAR-GP}) where $\Gamma_{0}\left(h\right)$
is given by Eq. (\ref{eq:RRFreeAssembly}). The idea here is to assess
the contribution to the SAR of the DC magnetic field through the dynamics,
or more precisely through the relaxation rate, and to compare it with
that brought in by the equilibrium susceptibility. Accordingly, we
see that the DC field $h_{{\rm max}}$, at which the SAR reaches its
maximum, is slightly shifted to higher values as the concentration
increases. However, the qualitative overall behavior remains the same.
This result implies that for the typical assemblies studied here,
the overall behavior of the SAR is mainly governed by the equilibrium
susceptibility $\chi^{{\rm eq}}$. There is a further remark in order
regarding the curves in Fig. \ref{fig:SAR-of-h} (b). At some particular
value of the DC applied field ($h_{c}\simeq0.07$), the SAR turns
out to be independent of the concentration. In fact, this occurs when
the DI contribution to the susceptibility ($\chi_{{\rm int}}^{{\rm eq}}$)
vanishes and thereby Eq. (\ref{eq:SAR-GP}) becomes independent of
$\tilde{\xi}$ (or the concentration). Hence, the exact expression
of $h_{c}$ can be obtained by solving $\chi_{\mathrm{int,GP}}^{\mathrm{eq}}\left(x_{c}\right)=0$,
or an approximation thereof at low field obtained from Eq. (\ref{eq:XiEqIntContr-1}),
leading to $h_{c}\sim\frac{1}{4\sigma}\left(1+\frac{1}{2\sigma}\right)\simeq0.05$.

\section{Discussions}

Now we discuss the main issue of the present work, namely the contributions
of DI and DC magnetic field to the SAR of an array of magnetic nanoparticles.
In particular, we discuss the role of the underlying super-lattice,
emphasizing its geometry and structure.

\subsection{Effect of the DC magnetic field on the SAR}

Both in the transverse and longitudinal static (DC) magnetic field
we find that the SAR exhibits a bell-like shape. However, in the transverse
setup the ascending part of the curve is rather abrupt and thus occurs
over a narrow range of low field values {[}see Fig. \ref{fig:Plot-of-the SAR in trans field}{]}.
This is probably the reason why only the descending part is observed
in experiments. For instance, in Ref. \onlinecite{kenya2013} the
authors studied the possibility to increase heating with the help
of a DC magnetic field in the transverse configuration. They measured
a SAR that is decreasing with increasing DC field, as can be seen
in their Fig. 4. Again, in Ref. \onlinecite{mehdaouietal12apl} the
authors studied the effect of a transverse DC magnetic field on the
SAR of FeCo nanoparticle assemblies. The results in their Fig. 3 (b)
show that the hysteresis area (or the SAR divided by the AC field
frequency) is a decreasing function of the DC field. However, if one
examines their results more closely in the low-field regime, it turns
out that a nonmonotonic behavior of the SAR is observed: close to
3 mT, it seems that the area of the hysteresis grows before decreasing
continuously. This is in agreement with the predictions of the present
theoretical developments. However, further experimental investigations
are required to clarify this behavior of the SAR. 

On the other hand,
the longitudinal setup renders a clearly nonmonotonic behavior of
the SAR with a maximum at a DC field that falls well within the experimental
range. The SAR can also be obtained by computing the area of the dynamic
hysteresis loop obtained by cycling over the AC magnetic field, see
Ref. \onlinecite{Carrey_JAP2011} and references therein. In the case
of an oblate geometry, the DI lead to an effective magnetic moment
in the plane of the assembly. Then, in the low-field regime, as we
increase the DC field, the projection of this magnetic moment (on
the field direction) increases, because the net magnetic moment tilts
out of the assembly plane, leading to a widening of the hysteresis
cycle $M(H_{AC})$ and thereby to an increase of the SAR. As the critical
value of the DC field is reached (see discussion above) the DC field
wins against the DI and the magnetization saturates. Consequently,
the equilibrium susceptibility and thereby $\chi^{\prime\prime}$
goes down to zero.

In Ref. \onlinecite{dejkal10j3m} the authors used the matrix-continued-fraction
method to study the effect of a DC magnetic field on the AC susceptibility
of a nanoparticle in the macrospin approximation. In particular, the
results in their Fig. 1 show that, in the low-frequency regime, which
is relevant to the present work, the imaginary component of the AC
susceptibility decreases as the DC field (denoted there by $\xi_{0}$)
is increased beyond unity. This is of course in agreement with the
behavior we observe here in the high-field regime, \emph{i.e.} $h>h_{c}$,
or $x>1$ as can be seen in Fig. \ref{fig:Linear-susceptibility-ksi=00003D0 and ksi non nul}
in the behavior of the equilibrium susceptibility. In fact, the latter
imposes its bell-like shape to the out-of-phase component of the AC
susceptibility and thereby to the SAR.

\subsection{Effects of the assembly super-lattice shape and structure\label{sec:Lattice-sums}}

In Eq. (\ref{eq:SuscepEq}) enters the lattice sum $\mathcal{C}^{\left(0,0\right)}$
leading to the effective DI parameter $\tilde{\xi}\equiv\xi\mathcal{C}^{\left(0,0\right)}$.
For the $2D$ array of $N$ particles considered here, $\mathcal{C}^{\left(0,0\right)}$is
defined by\citep{azzeggagh_EPJB2005}
\[
\mathcal{C}^{\left(0,0\right)}=\frac{1}{N}\sum_{i=1}^{N}\sum_{j\ne i}\frac{3\left(e_{ij}^{z}\right)^{2}-1}{r_{ij}^{3}},
\]
where $e_{ij}^{z}$ represents the $z$-component of $\bm{e}_{ij}$,
the unit vector introduced in Section \ref{sec:Model-and-Hypotheses}.
\begin{figure}[h]
\begin{centering}
\includegraphics[width=1\columnwidth]{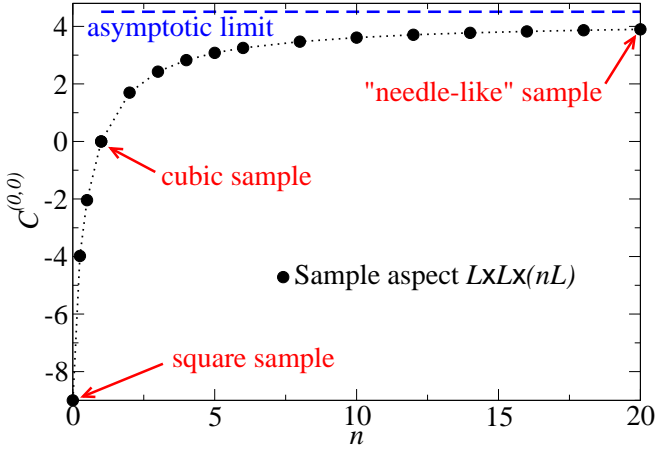}
\par\end{centering}

\caption{Systematic evaluation of the lattice sum $\mathcal{C}^{\left(0,0\right)}$
as a function of $n$ for $N$-particle samples with $N=L\times L\times nL$,
note that $n=0$ means a 2D square sample.\label{fig:C(0,0)}}
\end{figure}
 These lattice sums can easily be computed on the simple cubic lattice
for samples with different shapes, with $L_{x}=L;\ L_{y}=L;\ L_{z}=nL$,
by varying $n$. Namely, $n=0$ corresponds to a square sample, $n=1$
to a cubic sample, and if $n\gg1$ the sample assumes the shape of
a needle. The results of the evaluation of $\mathcal{C}^{\left(0,0\right)}$
at the thermodynamic limit are given in Fig. \ref{fig:C(0,0)} and
are summarized in the following Table: 

\begin{center}
\begin{tabular}{|c|c|c|c|c|}
\cline{1-2} \cline{4-5} 
{$n$} & {$\mathcal{C}^{\left(0,0\right)}\approx$} &  & {$n$} & {$\mathcal{C}^{\left(0,0\right)}\approx$}\tabularnewline
\cline{1-2} \cline{4-5} 
{0} & {-9.1} &  & {6} & {3.25}\tabularnewline
\cline{1-2} \cline{4-5} 
{1/4} & {-3.98} &  & {8} & {3.47}\tabularnewline
\cline{1-2} \cline{4-5} 
{1/2} & {-2.04} &  & {10} & {3.61}\tabularnewline
\cline{1-2} \cline{4-5} 
{1} & {0} &  & {12} & {3.70}\tabularnewline
\cline{1-2} \cline{4-5} 
{2} & {1.69} &  & {14} & {3.77}\tabularnewline
\cline{1-2} \cline{4-5} 
{3} & {2.42} &  & {16} & {3.82}\tabularnewline
\cline{1-2} \cline{4-5} 
{4} & {2.82} &  & {18} & {3.86}\tabularnewline
\cline{1-2} \cline{4-5} 
{5} & {3.08} &  & {20} & {3.89}\tabularnewline
\cline{1-2} \cline{4-5} 
\end{tabular}
\par\end{center}{\par}

In order to highlight the competition between the applied field and
the DI, we mainly focus on oblate samples. This implies that $\mathcal{C}^{\left(0,0\right)}<0$
and as it can be seen in Eq. (\ref{eq:SuscepEq}), the sample aspect
ratio plays a key role in the behavior of $\chi^{{\rm eq}}$: while
the role of the interactions is irrelevant for cubic samples since
$\mathcal{C}^{\left(0,0\right)}=0$, it gets more and more enhanced
as one goes to planar samples as observed in Fig. \ref{fig:Chi_f=00005BC(0,0)=00005D}. 

\begin{figure}[H]
\begin{centering}
\includegraphics[width=1\columnwidth]{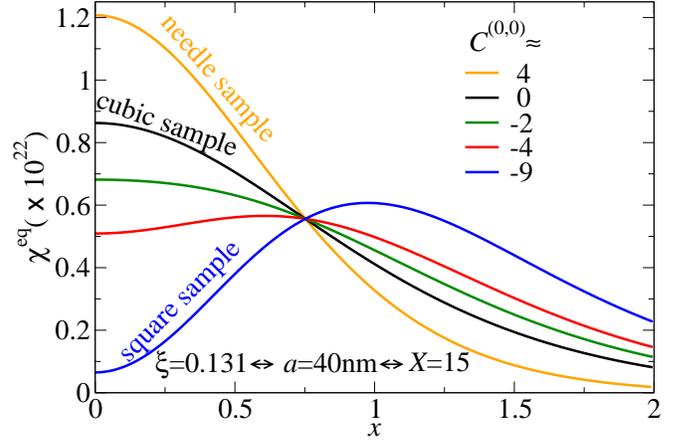}
\par\end{centering}

\caption{Equilibrium susceptibility given by Eqs. (\ref{eq:SuscepEq}), (\ref{eq:Xifree GP})
and (\ref{eq:Xi int GP}) as a function of the DC field for samples with different aspect ratios:
$L\times L\times 20L$ (orange), $L\times L\times L$ (black), $L\times L\times\frac{L}{2}$ (green),
$L\times L\times\frac{L}{4}$ (red), and $L\times L$ (blue).\label{fig:Chi_f=00005BC(0,0)=00005D}}
\end{figure}

As a consequence, the SAR changes from a bell-like curve (with a maximum)
into a monotonously decreasing function of the applied DC field, as
the sample passes from a pure $2D$ array into a thick slab and finally a
``needle-like'' sample. This
means that the assembly shape plays a crucial role in the present
approach. For the case of prolate samples, with
$\mathcal{C}^{\left(0,0\right)}>0$, the DC field does not compete with the DI,
and both have the same effect on the anisotropy barrier, this translates into
a decrease of the SAR as the DC field increases. 
In contrast, for the well controlled organized $2D$ arrays
of nano-elements in vogue today (with $\mathcal{C}^{\left(0,0\right)}<0$), 
there is a competition between the DI and the
DC field, such that the SAR should exhibit the bell-like shape.

\section{Conclusion and Perspectives}

We have proposed a theoretical model and a practical tool for studying
the qualitative behavior of the specific absorption rate of a monodisperse
assembly of magnetic nanoparticles with oriented effective anisotropy,
in the presence of dipolar interactions and a DC magnetic field, in
addition of course to the AC field. We have dealt with both a longitudinal
and transverse setup of the DC field with respect to the anisotropy
axis. We have shown that, depending on the sample geometry, one can
observe competing effects between the external DC field and the sample's
concentration (or equivalently the dipolar interaction). More precisely,
in the case of oblate samples and for a given concentration, there
is an optimal field magnitude that maximizes the specific absorption
rate. 

In the present work, we have modeled the magnetic state of the nanoparticles
with the help of a macroscopic magnetic moment, thus ignoring their
internal structure and intrinsic features, such as surface effects.
In Ref. \onlinecite{Kachkachi:2002aa} the effect of surface anisotropy
on the (static) hysteresis loop was investigated in the atomic approach
with the help of numerical methods. The same approach could be used
to compute the dynamic hysteresis loops and thereby investigate the
effect of surface and finite size on the specific absorption rate.
The corresponding results could also be compared with the approach
used here that uses the AC susceptibility upon extending it to the
effective one-spin problem along the lines adopted in Ref. \onlinecite{Vernay_etal_acsucept_PRB2014}.

As mentioned earlier, the nonlinear regime with respect to the AC
magnetic field should be studied upon generalizing the Debye model
for AC susceptibility. An increase in the AC field affects the magnitude
of the SAR explicitly through the prefactor and implicitly through
the relaxation rate and the additional contributions from high frequency
modes.
\begin{acknowledgments}
We would like to acknowledge instructive discussions with Oksana Chubykalo-Fesenko,
David Serantes and Carlos Boubeta. 
\end{acknowledgments}

\bibliography{ma_biblio}

\end{document}